\documentclass[prc,twocolumn,secnumroman,tightenlines,amssymb,aps,nobibnotes,superscriptaddress,balancelastpage,floatfix,nofootinbib]{revtex4}
\usepackage{soul}  % \textst{..} -> skreslenie

\usepackage{amsmath}
\usepackage{slashbox}
\usepackage{soul}
\usepackage{float}
\usepackage{lipsum}
\usepackage{graphicx}
\usepackage{multirow}           
\usepackage{color}
\DeclareRobustCommand\substyle{\name@idx{document substyle}}
\DeclareRobustCommand\classoption{\name@idx{document class option}}
\DeclareRobustCommand\classname{\name@idx{document class}}
\def\name@idx#1#2{{\ttfamily#2}
\index{#2\space#1=\string\ttt{#2}\space#1}\index{#1>#2=\string\ttt{#2}}}
\DeclareTextFontCommand{\rb}{\color{red}\bfseries}

\begin{document}

\title{Collectivity of rotational motion in $^{220}$Rn and $^{226}$Ra}

\author{A. Dobrowolski}
\email{arturd@kft.umcs.lublin.pl}
\affiliation {Department of Theoretical Physics, University of Maria Curie-Sk\l{}odowska, pl. Marii Curie-Sk\l{}odowskiej 1, PL-20031 Lublin, Poland}
\author{K. Mazurek}
\email{Katarzyna.Mazurek@ifj.edu.pl}
\affiliation{Institute of Nuclear Physics PAN, ul.\,Radzikowskiego 152,Pl-31342 Krak\'ow, Poland}

\author{K. Pomorski}
\email{Krzysztof.Pomorski@ncbj.gov.pl}
\affiliation{National Centre for Nuclear Research, ul. Pasteura 7, 02-093 Warsaw, Poland}
\date{\today}

\begin{abstract}  
Calculations to reconstruct rotational level patterns in the $^{220}$Rn and $^{226}$Ra nuclei have been performed using a collective quadrupole+octupole approach with microscopic mass tensor and moments of inertia dependent on deformation and pairing degrees of freedom. The main objective is to quantitatively confirm the known experimental observations that the Rn nucleus passes from octupole vibrational to octupole deformed with increasing rotation frequency, while the Ra nucleus is relatively weakly affected by collective rotation, being octupole deformed from the beginning.
The collective potential in a nine-dimensional collective space is determined using the macroscopic-microscopic method with Strutinsky and the BCS with an approximate particle number projection microscopic corrections. The corresponding Hamiltonian is diagonalized based on the projected solutions of the harmonic oscillators coupled with Wigner functions. Such an orthogonalized basis is additionally symmetrized with respect to the so-called intrinsic symmetrization group, specifically dedicated to the collective space used, to ensure the uniqueness of the Hamiltonian eigen-solutions in the laboratory frame.
The response of the pairing and deformation degrees of freedom to external rotation is discussed in the variational approach, where the total energy is minimized by the deformation and pairing variables. As the nuclear spin increases, the pairing gaps of protons and neutrons decrease from its ground-state equilibrium values to zero (no superfluid solution). 
Consequently, the corresponding microscopic moments of inertia increase with collective spin (Coriolis {\it antiparing} effect), resulting in effectively lower rotational energy levels I$^{\pi}$ with respect to pure classical-rotor pattern I(I+1). The expression for {\it cranking} microscopic moments of inertia allows to discuss the rotational Hamiltonian term as depending on the nuclear structure. The effect of vibrations of the pairing field is considered only on average by introducing a multiplicative constant factor to rescale the moment of inertia values.
The obtained comparison of experimental and theoretical rotational energy level schemes, dipole, quadrupole and octupole transition probabilities of B(E$\lambda$) in $^{220}$Rn and $^{226}$Ra is satisfactory.

\end{abstract}

%For fun, see PACS list at  https://www.aip.org/publishing/pacs/pacs-2010-regular-edition

\maketitle 

\section{Introduction}
Spectroscopic observations of collective bands, especially of negative parity, carried out since roughly the middle of the last century, allow us to distinguish two main types of octupole instabilities in atomic nuclei: (a) a vibration of a quadrupole-deformed nucleus surface, breaking the reflection symmetry with respect to the XOY plane of the body-fixed frame while preserving the axial symmetry, alternatively, (b) the possession by the nucleus of a more or less permanent octupole deformation \cite{butler,cocks}. The latter scenario is realized in nuclei around the so-called octupole magic numbers $Z, N=34, 56, 88, 134,...$, that is, at values just greater than the magic numbers where the nuclei are nearly spherical. In such nuclei, the appearance of octupole deformation entails the appearance of intruder states close to the Fermi level, allowing for octupole-type interaction $~Y_{30}$ with the occupied particle states. The magnitude of this coupling, depending mainly on the difference in their energies and single-particle angular momenta, which must be $\Delta l=\Delta j=3$, determines essentially the depth of the double-degenerated octupole minima located symmetrically around $\beta_3=0$ and therefore also the persistence of such a configuration due to collective rotation.

Until recently, it was thought that breaking the intrinsic axial symmetry (defined in the intrinsic reference system) by the octupole interaction seems to be very unlikely. Nevertheless, in the works
\cite{Dudek-Dedes1,Dudek-Dedes2} the existence of energy states with characteristic degrees of their degeneracy that have not been considered so far in nuclear spectroscopy is seriously discussed. It is known from elementary group theory that this degree of degeneracy is directly related to the intrinsic symmetry of the nucleus and the dimensions of its irreducible representations. However, the symmetry most commonly discussed in the context of octupole instabilities remains the axial symmetry created only by pear-like shapes proportional to $Y_{20}+Y_{30}$ harmonics. Few collective models treat the deformations of the quadrupole and octupole with effective triaxiality, that is, \cite{nadirbekov} with rotation and vibration included or the PROXY-SU(3) symmetry model \cite{proxy}. An overview of the results for various collective models is presented in \cite{bonatsos}.

Considering possible non-axial octupole degrees of freedom, we open up the opportunity to discuss the occurrence of states or even entire bands with non-axial, breaking the XOY reflection symmetries, often referred to as {\it exotic symmetries}. Even if the aspect of new exotic symmetries is not explicitly addressed in this work, the model from which the results of \cite{AD-KM-AG-2016,AD-KM-AG-2018,AD-AG-KM-2018} were obtained allows the description of rotational states built on the full set of six, axial and non-axial quadrupole and octupole collective excitations and also the hexadecapole parameter $\alpha_{40}$ with respect to which total potential energy is minimized. All of these describe the nuclear surface in the body-fixed coordinate system.

In order to ensure the unambiguousness of the nuclear wave functions in the laboratory reference system in which the nucleus is measured, we apply a symmetrization procedure to obtain eigenstates of the Hamiltonian, acting in the space of intrinsic variables (deformations), with respect to a specific symmetrization group, different, however, from that octahedral group for the well-known Bohr Hamiltonian \cite{BohrHam} in quadrupole variables. The results presented in the following sections are obtained within the framework of this model, which has, however, undergone significant modifications with respect to the original version mentioned, improving the ability to reproduce the level energies, the intraband and interband B(E$\lambda$) transitions compared to less complex rigid-rotor nuclei.

In the present work, our aim is to generate the low-lying rotational even-spin ground-state band (GSB) with positive parity, and a low-lying odd-spin negative-parity band, as well as related to the GSB through B(E1) dipole and B(E3) octupole electric transitions for the $^{226}$Ra and $^{220}$Rn nuclei. In the work \cite{cocks}, the heavier nucleus is considered to have a permanent axial-octupole deformation while the lighter one is supposed to be vibrational. The vibrational nature of Radon isotope gradually transforms into an octupole-deformed nucleus as the collective rotation frequency increases, starting from its lowest values.
Such characteristics of the studied nuclei can be evidenced by experimental spin alignments as a function of rotation frequency, which are extensively reported in the above articles, and also by the energy interval between the states $0^+$ and $1^-$ (or $3^-$). In the energy spectrum of $^{220}$Rn, this substantial interval of around 0.64 MeV is explained by a coupling of collective rotation with an octupole phonon with the angular momentum of $~3\hbar$ performing precession around the OZ axis, tending to align quickly with the rotation axis as the rotation frequency increases. 

The rotation of the well-deformed octupole configuration observed in the $^{226}$Ra nucleus, being the effect of the aforementioned significant coupling of the valence single-particle state to an intruder of opposite parity with $\Delta j=\Delta l=3$, the $1^-$ state built on it lies higher than $0^+$ of the GSB only by $253$ keV (compared to $645$ keV in $^{220}$Rn). This energy difference is, of course, related to the purely rotational excitation and, in addition, to the difference in the zero-point vibrational levels of the octupole and quadrupole phonons, which are the vibrational heads of both bands.

The Sec.~\ref{theory} a short description of the applied theoretical models is presented, followed by results (Sec.~\ref{results}) for moments of inertia and pairing interaction, which give the total energy estimates. The transition probabilities are compared with experimental data and concluded in Sec.~\ref{wnioski}.

\section{Theory}\label{theory}
%P1: Intro: scientific motivation
\subsection{Modeling collective vibrations and rotations}
\label{shape_par}
The basic means to describe vibrational motion of the nuclear surface is to expand it in the intrinsic reference frame in terms of the orthogonal set of the spherical harmonics $\{Y_{\lambda\mu}(\vartheta,\varphi)\}$ expressed in spherical coordinates. The use of such type of shape parameterization given by, 
\begin{equation}
      R(\vartheta,\varphi)= 
      R_{0}c(\alpha)
      \big[ 
           1 + \sum_{\lambda=1}^{3} 
               \sum_{\mu=-\lambda}^{\lambda} 
               (\alpha_{\lambda \mu}^{\star})
                    Y_{\lambda \mu}^{} (\vartheta,\varphi),
     \big],	                                                            
\label{eqn.1}
\end{equation}
limited to the dipole ($\lambda=1$), quadrupole ($\lambda=2$) and octupole ($\lambda=3$) terms offers for the present study the exact treatment of the uniqueness of the intrinsic Hamiltonian eigensolutions in the laboratory reference frame and allows to calculate the B(E$\lambda$) transition rates in a systematic way, taking into account the couplings between all the above mentioned multipole vibrational modes.

The function $c(\alpha)$ depending on all $\alpha$-type variables is obtained from the volume conservation condition. The dipole $\alpha_{10}$ and $\alpha_{1\pm 1}$
deformation parameters are determined from the condition that the center of mass of the nuclear body, for any value of the quadrupole and octupole deformation, is shifted to the onset of the coordinate system and thus are dependent on $\{\alpha_{2\nu}, \alpha_{3\mu}\}$ independent variables.
Although dipole coordinates $\alpha_{1\nu}$ are not independent within this model, their values calculated as described above are taken into account in electric multipole transition operators, widely discussed in 
\cite{AD-KM-AG-2016}.

For a spherical tensor defined in $SO(3)$ manifold, represented e.g. by $\alpha_{\lambda\mu}$ or 
$Y_{\lambda\mu}$, the relation
$\alpha_{\lambda\mu}^{\star}=(-1)^{\nu}\alpha_{\lambda-\mu}$ obviously holds true.
As the space of effectively two quadrupole variables ($\alpha_{20}$,  $\alpha_{22}=\alpha_{2-2}$) with vanishing $\alpha_{21}=\alpha_{2-1}=0$
determines the direction of a body-fixed frame axes with respect to the nuclear surface, the full octupole 
$\{\alpha_{3\nu}\}, \nu=0,1,2,3$ complex tensor together with Euler angles form the twelve-dimensional collective-variable space. Using, in addition, 
often applied conditions that the imaginary parts of the $\alpha_{3\mu}$ tensor vanish, i.e.
 ${\rm Im}(\alpha_{3\mu})=0$, we are end up with nine collective real variables to define our Hamiltonian and the resulting wave functions. Although such a condition limits the full class of octupole shapes, it nevertheless preserves shapes that are physically relevant, especially in the context of above mentioned exotic symmetries. 
Mention, that due to the fact that the low-lying rotational bands in question are built on normal deformed configurations with small and moderate deformations of $(\beta_2,\beta_3)$, so there is no concern that the formula (\ref{eqn.1}) will produce completely nonphysical shapes, as could be the case, for example, when discussing the nuclear fission process. 

\subsection{Collective Hamiltonian}

The vibrational-rotational Hamiltonian defined in the set of $\alpha_{\lambda \mu}$ variables,
reproducing the shape of the nuclear system in the body-fixed frame, and the three Euler angles
$\Omega$, giving its orientation with respect to the laboratory frame, seem to produce a reasonable approximation of measured spectroscopic observables, such as rotational energies and the B(E$\lambda$) transition probabilities. A further simplification of this approach may rely on the application of the so-called {\it adiabatic approximation} that leads to a separation of the vibrational and rotational motions. Such separation may be possible because of the significantly different energy scales of both of these collective modes.
Neglecting the coupling between the collective-variable subspaces $\{\alpha_{2\nu}\}$ and $\{\alpha_{3\mu}\}$, the two independent mass tensors for the vibrational quadrupole and octupole modes are determined.
Now, profiting of the above approximations, a realistic, quantized quadrupole-octupole Hamiltonian
with varying mass parameters and microscopic {\it cranking} moments of inertia \cite{Inglis} can be written, as e.g. in \cite{AD-KM-AG-2016,AD-KM-AG-2018}

 \begin{eqnarray}
 && {\mathcal H}_{coll}(\alpha_{\textbf{2}},\alpha_\textbf{3},\Omega)=\frac{-\hbar^2}{2}\bigg\{\nonumber\\
 && \frac{1}{\sqrt{\vert B_2\vert}}\sum\limits_{\nu\nu^{\prime}=0}^2
  \frac{\partial}{\partial \alpha_{2\nu}}
   \sqrt{\vert B_2\vert} \big[B_2^{-1}\big]^{\nu\nu^{\prime}}
    \!\!\!\frac{\partial\;\;}{\partial \alpha_{2\nu^{\prime}}}+\nonumber\\
 &&  \frac{1}{\sqrt{|B_3|}}\sum\limits_{\mu\mu^{\prime}=0}^3
   \frac{\partial}{\partial \alpha_{3\mu}}
   \sqrt{|B_3|} \big[B_3^{-1}\big]^{\mu\mu^{\prime}}
   \!\!\!\frac{\partial\;\;}{\partial \alpha_{3\mu^{\prime}}}\bigg\}+ \nonumber\\
&& \hat H_{rot}(\Omega) + \hat V(\alpha_\textbf{2},\alpha_\textbf{3})
\label{Hcoll}                                                   %\label{eqn.02}
\end{eqnarray}
where $\alpha_\textbf{2}$ and $\alpha_\textbf{3}$ denote symbolically the quadrupole and octupole subspaces, respectively, while $B_{2}(\alpha_\textbf{2})$ and $B_{3}(\alpha_\textbf{3})$ are the corresponding, widely used in the nuclear structure physics microscopic {\it cranking} mass tensors \cite{Inglis,crank} with their determinants
$|B_2|={\mathrm det}B_{2\mu}(\alpha_{2\mu})$ and $|B_3|={\mathrm det}B_{3\mu}(\alpha_{3\mu})$. 

In an explicit form, $(B_{\lambda})_{\nu\nu'}(\alpha_{\lambda})$ reads:
\begin{eqnarray}
(B_{\lambda})_{\nu\nu'}=\sum\limits_{kl}
\frac{\langle k|\frac{\partial\hat H_{sp}}{\partial\alpha_{\lambda\nu}} |l\rangle\, 
      \langle l|\frac{\partial\hat H_{sp}}{\partial\alpha_{\lambda\nu'}}|k\rangle}
        {(E_k+E_l)^3} \big(u_k\,v_l+v_k\,u_l\big)^2,
\label{cran_mass}\end{eqnarray}

where the double summation runs over the full set of the BCS quasi-particles, obtained out of the eigensolutions of the folded-Yukawa mean-field
Hamiltonian $\hat H_{sp}$ of Ref.~\cite{myCPC2016} and used pairing approach. The amplitudes of the occupation probability $v_n$ and $u_n$ are related by $u_n^2=1-v_n^2$ and $E_n$ is the energy of the $n^{th}$ quasiparticle.

Nowadays, the potential energy of a nucleus is usually generated within various self-consistent microscopic models, however, in this work, due to the very extensive six-dimensional space of nuclear deformations, a more effective solution seems to be reasonable. Thus, calculations are performed with a still widely exploited macroscopic-microscopic model with Strutinsky shell correction of $6^{th}$ order and particle-projected BCS approach \cite{colpair,stab,pomorski1}. The strength of the pairing force is given by an (A,Z)-dependent phenomenological formula of Ref.~\cite{dudek80}. For a wise choice of mean-field potential, pairing interaction and the smooth liquid-drop energy prescription, the macroscopic-microscopic method offers quite realistic estimates of $\hat V(\alpha_\textbf{2},\alpha_\textbf{3})$ function. 

Due to significantly different energy scales of the collective vibrational and rotational motions, they are assumed here to be totally decoupled. Consequently, the rotational Hamiltonian term
$\hat{H_{rot}}$ depends only on the Euler angles $\Omega$, and parametrically on the static
deformation and the pairing gap parameters $(\Delta_n,\Delta_p)$ for protons and neutrons,
determined in equilibrium state by a given spin value $I$. The spin-dependent energy equilibrium appears when the potential plus the rotational energy of a rotating nucleus of given spin in the form:
\begin{eqnarray}
E_{tot}^I(\alpha_\textbf{2},\alpha_\textbf{3},\Delta_n,\Delta_p;&&\!\!\!\!\!\! I)=
V(\alpha_\textbf{2},\alpha_\textbf{3},\Delta_n,\Delta_p)+ \nonumber\\
&&\frac{I(I+1)\hbar^2}{2J_{\perp}(\alpha_\textbf{2},\alpha_\textbf{3},\Delta_n,\Delta_p)}
\label{Etot}\end{eqnarray}
reaches its minimum in the deformation-pairing space. 
Obviously, the cranking moment of inertia with respect to the axis perpendicular to the symmetry OZ axis,  $J_{\perp}(\alpha_\textbf{2},\alpha_\textbf{3},\Delta_n,\Delta_p)$, entering Eq.~(\ref{Etot}), is calculated for a given spin value $I$, the actual deformation and the BCS occupation probability amplitudes $u$ and $v$, as:
\begin{equation}
J_{\perp}(\alpha_\textbf{2},\alpha_\textbf{3},\Delta_n,\Delta_p)=2\hbar^2\sum\limits_{kl}\frac{|\langle k | 
\hat j_{\perp}|l\rangle|^2}{E_{k}+E_{l}}(u_{k}v_{l}-u_{l}v_{k})^2.
\label{crankingMoI}\end{equation}
 The operator $\hat j_{\perp}$ expresses the single-particle total angular momentum along the axis perpendicular to OZ. If the non-axiality $\alpha_{22}$ is taken into account, $J_{\perp}$ can be either $J_x$ or $J_y$, depending which of them is larger. This choice, similarly as in classical mechanics, guarantees the strongest stability of the collective rotation.

Since, as widely discussed in \cite{AD-KM-AG-2016, AD-KM-AG-2018}, the eigenstates of the Hamiltonian 
(\ref{Hcoll}) must be symmetrized with respect to the specific symmetrization 
$\bar{G_s}=D_{4y}$ group, the Hamiltonian itself, (including rotational Hamiltonian term) has to be scalar with respect to this group. Then, by special choice of the deformation space, the potential energy operator 
$\hat V(\alpha_\textbf{2},\alpha_\textbf{3},\Delta_n,\Delta_p)$ and the kinetic term are both $D_{4y}-$symmetric by construction. An identical property must occur for the rotor Hamiltonian.
In contrast to the traditional form of the triaxial, $\bar{D_2}-$symmetric rotor, it is convenient to construct it out of the irreducible tensors of $\bar{G_s}$ group, as presented in details in 
Refs.~\cite{gen_Hrot,gen_Hrot2,AD-KM-AG-2016,AD-KM-AG-2018}. 
In this context, one can say about a systematic way of building rotor Hamiltonians, symmetric with respect to a demanded symmetry group. Let us emphasize that $\bar{G_s}$ should be considered as a subgroup of the full symmetry group of the Hamiltonian associated only with purely rotational transformations, which
can, in general, be different if other than mentioned in subsection (\ref{shape_par}) set of 
$\alpha_{\lambda\mu}$ shape variables is chosen.

Having diagonalized the Hamiltonian (\ref{Hcoll}) in the space of $D_{4y}$ symmetrized basis functions
constructed out of the six-dimensional harmonic-oscillator multiphonon solutions, as in Refs. \cite{AD-KM-AG-2016,AD-KM-AG-2018}, we obtain full set of collective energies for which the reduced probabilities of electric dipole, quadrupole, and octupole transitions B(E$\lambda$) are determined.
A certain subset of its low-energy eigenstates, which have collective energies together with the interband and intraband B(E$\lambda$) values close to those, measured for low-lying bands, can be considered as the models of experimentally populated spectra of given spin, parity etc.

\section{Results}\label{results}
\subsection{Total energy}
 The shape evolution of the nucleus with an increase in angular momentum is still a complex problem. The open question is whether the nuclear shape can change within a single band and if there exists a possibility to notice it. It is clear that in this context, the presence of microscopic shell effects whose energies are a fraction of the value of the nucleus' liquid-drop energy, are crucial for low-energy rotational transitions of the order of tens of keV, considered in this work.

In the present model, based on the intrinsic configurations characterized by the dipole, quadrupole and octupole deformation tensors, ${\alpha_{\lambda\mu}}$, or the corresponding multipole moments ${Q_{\lambda\mu}}$, Euler angles and parameters describing the field of pairing interactions, 
 $\Delta_{p(n)}$, the rotational states are obtained as eigen-solutions of the full vibration-rotation Hamiltonian (\ref{Hcoll}).

 The pairing interaction, determined in the BCS method with an approximate projection onto a good number of particles \cite{stab}, are included both in the potential energy function of the macroscopic-microscopic model in the form of a pairing energy correction to the LSD liquid drop energy \cite{Pomorski-Dudek-2003} and through the amplitudes of the pairing probabilities in the microscopic cranking formula \cite{Inglis} on the moments of inertia. The idea of using the moments of inertia varying with nuclear spin in the rotational part of the Hamiltonian gives, in a macroscopic way, the influence of Coriolis and centrifugal forces on the energies and structure of rotational states.
 
Currently, the total energy, for a given spin value, is minimized over axial and non-axial quadrupole  
($\alpha_{20},\alpha_{22}$), octupole ($\alpha_{30}$,$\alpha_{31},\alpha_{32},\alpha_{33}$) and hexadecapole ($\alpha_{40}$) deformation parameters. Additionally, the minimization  has also been performed over the pairing $\Delta_{n,p}$ for neutrons and protons, independently. 

\begin{figure}[H]
\includegraphics[width=3.0in,angle=0]{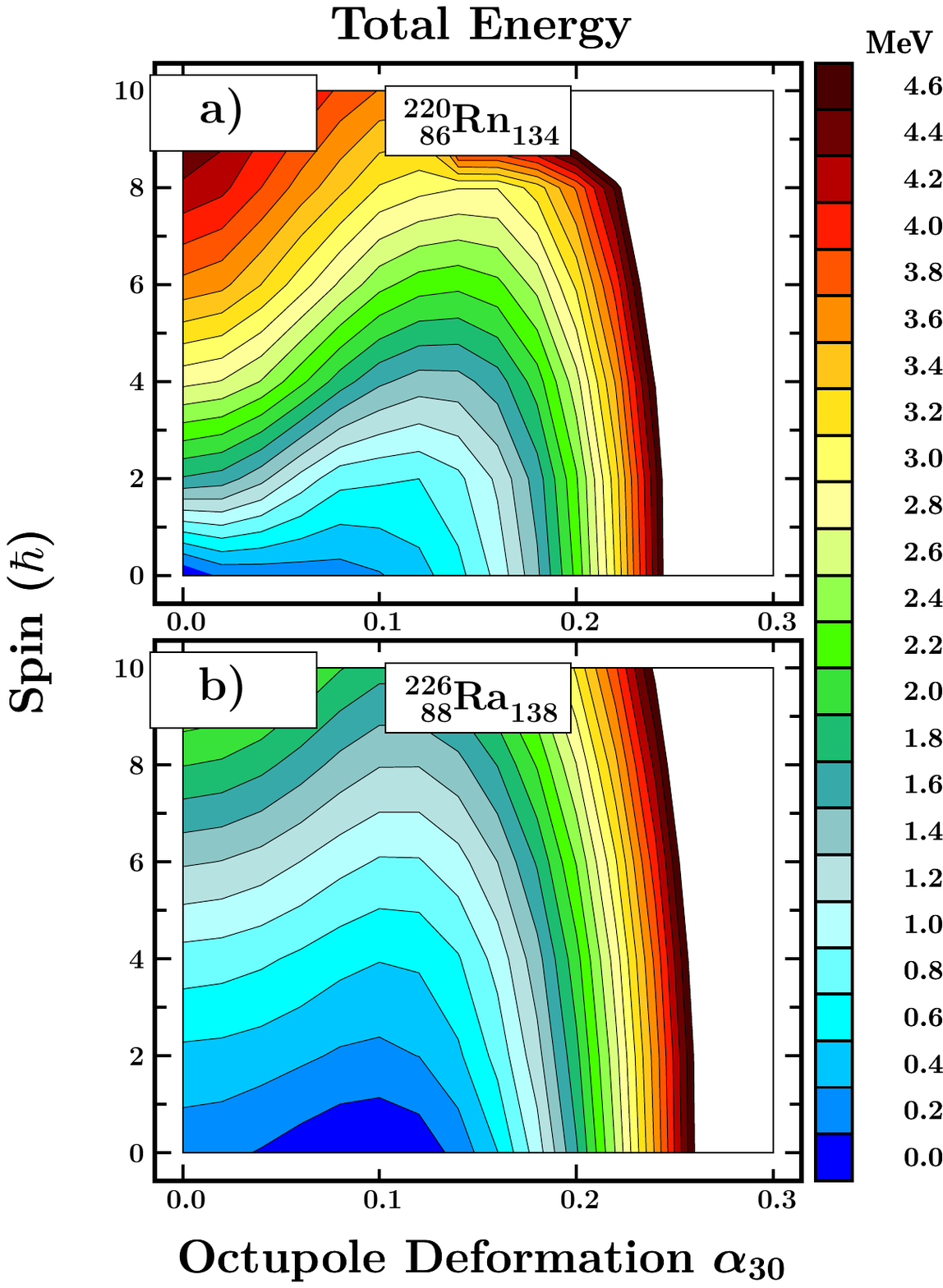}
\caption{\label{fig03} The total energy minimized over deformation parameters ($\alpha_{20},\alpha_{22},\alpha_{3\nu}, \nu=1,2,3$) and pairing $\Delta_{n}$, $\Delta_{p}$ gap parameters, displayed in spin $I$ and axial-octupole deformation plane for $^{220}$Rn and $^{226}$Ra.
}
\end{figure}
The global behavior of the energy minimum in the spin-mass asymmetry plane is presented in Fig.~\ref{fig03}.
The minimum of total energy changes smoothly towards higher octupole deformation and is well localized for the almost constant value of $\alpha_{30}\approx 0.17$ for $^{226}$Ra. For lighter $^{220}$Rn nucleus, this trend is more dynamical leading to a rapid energy increase as function of spin. In addition to the axial deformation parameters $\alpha_{20}$ and $\alpha_{30}$, other non-axial degrees of freedom describing the quadrupole and octupole shape oscillations of the nuclei in question are not excited in low spin collective rotational motion.

\begin{figure}[!bt]
\setlength{\unitlength}{0.1\textwidth}
\begin{picture}(3,7.5)
%\put(2,9){a)} 
\put(-0.9,3.4){\includegraphics[width=3.0in]{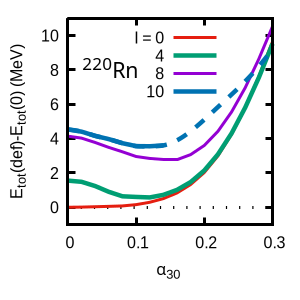}}
\put(2.6,5.){\bf \large a)} 
    \put(-0.9,0){\includegraphics[width=3.0in]{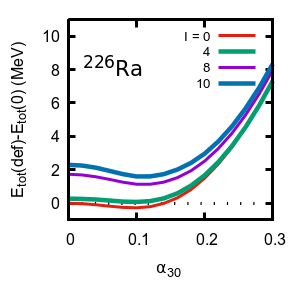}}
\put(2.6,1.5){\bf \large b)}
\end{picture}
\caption{\label{profil} Evolution of the total energy for $^{220}$Rn (a) and $^{226}$Ra (b) with spin.
}\label{fig01}
\end{figure}

A more convincing comparison of energy profiles is shown in Figs.~\ref{fig01}, where the lines go through the minima of total energy (\ref{Etot}) with respect to the energy of the spherical nucleus for given spins, $I=0, 4, 8, 10~\hbar$ for (a) $^{220}$Rn and (b) $^{226}$Ra as a function of the octupole deformation parameter. The energy lines in $^{226}$Ra corresponding to different spins are almost parallel, revealing very little effect of collective rotation on the shape of, in fact, quite shallow octupole energy well, which exists even in a 
non-rotating system. A tiny depth of the octupole well of around 0.5 MeV in $^{226}$Ra is also reported in other theoretical estimates, e.g. \cite{butler}.
As also predicted in \cite{cocks}, this nucleus does not undergo significant rearrangements in the intrinsic structure; however, the variable field of neutron pairing indicates that this nucleus is not at all a classcal rotor. This can also be easily observed by examining the distortions of the spacing between measured levels from the $I(I+1)$ pattern in the low-lying bands, also shown in \cite{cocks}.

 The case of $^{220}$Rn is more interesting as the location and depth of the minimum octupole energy change abruptly with spin. As mentioned earlier, due to the relatively large softness of the potential energy surface around the ground-state point along axial octupole shapes and mainly neutron pairing-induced rapid changes in microscopic moments of inertia as a function of deformation and spin, the increasing rotation rate provokes an immediate shift of the equilibrium state towards increasingly reflection-asymmetric shapes.
 
%%%%%%%%%%%%%%%%%%%%%%%%%%%%%%%%%%%%%%%%%%%%
\subsection{Impact of pairing correlations}
%%%%%%%%%%%%%%%%%%%%%%%%%%%%%%%%%%%%%%%%%%%%%

\begin{figure}[!bt]
\setlength{\unitlength}{0.1\textwidth}
\begin{picture}(3,7.5)    
\put(-0.9,3.5){\includegraphics[width=3.50in]{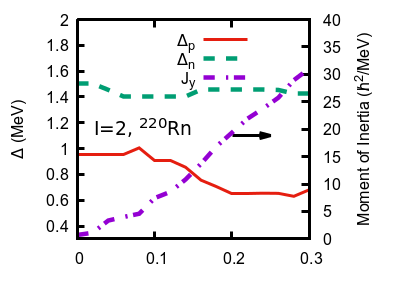}}
\put(0.2,6.5){\bf \large a)} 
\put(-0.9,0){\includegraphics[width=3.50in]{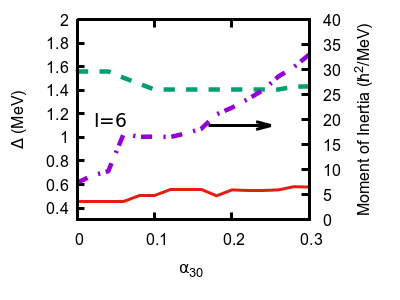}}
\put(0.2,3.){\bf \large b)}
\end{picture}
\caption{Pairing energy gap $\Delta$ values for protons (red) and neutrons (green) obtained after energy minimization for selected spin values, $I=2~\hbar$ (a) and 6~$\hbar$ (b) for \label{fig05} in $^{220}$Rn. The evolution of moment of inertia value, ($J_{\perp}$), is also displayed (magenta).\label{fig05} 
}
\end{figure}
%%%%%%%%%%%
%\begin{widetext}
\begin{figure*}[!bt]
\includegraphics[width=3.in]{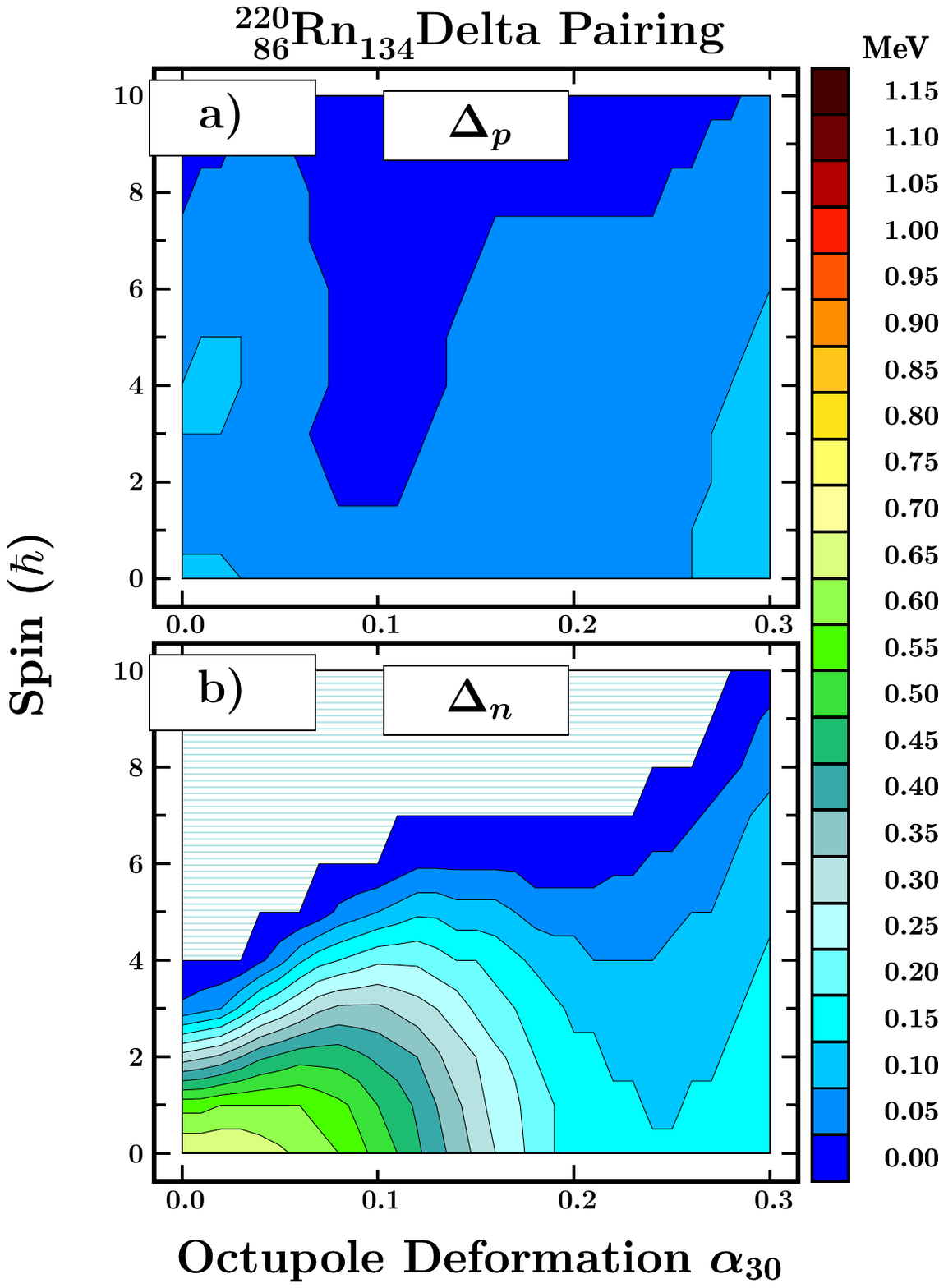}
\includegraphics[width=3.in]{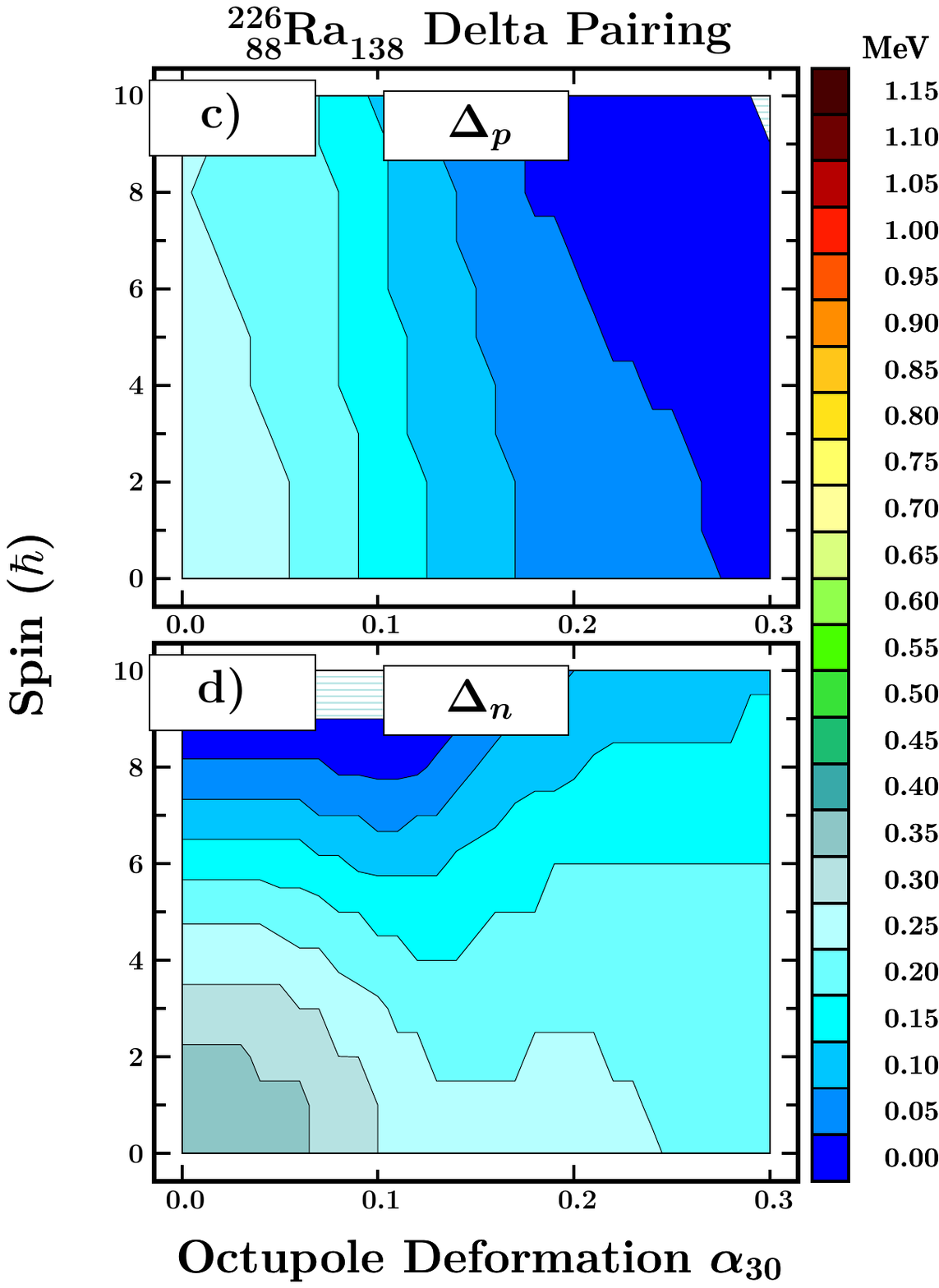}
\caption{\label{delty} Proton and neutron pairing $\Delta$ values in spin and deformation $\alpha_{30}$ plane
compared for $^{220}$Rn (a,b) and $^{226}$Ra (c,d), obtained by minimizing the total energy (\ref{Etot}) over all axial and non-axial quadrupole and octupole deformation parameters.
}

\end{figure*}
%\end{widetext}
The pairing correlations are known to be sensitive to collective processes. In particular, the increase of the velocity of collective rotation causes, in general, a weakening of the pairing mean-field. The critical spin where pairing abruptly decreases due to Cooper pair scattering depends on the pattern of single particle levels around Fermi level, as e.g. density of levels or presence of energy gaps which may change with deformation, excitation energy etc. The behavior of pairing degrees of freedom in a given collective process can be reconstructed in a natural way in microscopic theories, as e.g. commonly used Hartree-Fock-Bogoliubov or a more simplistic version of this known as Hartree-Fock-Bogoliubov+Cranking (HFBC) developed, e.g. in \cite{devoight-dudek}. However, a similar dependence of the pairing $\Delta$ in the equilibrium point for a given spin $I$ can also be reproduced within a variational scheme by minimizing the total macroscopic-microscopic energy with an additional classical rotational term dependent on the deformation and the pairing-dependent microscopic moment of inertia, as used in Eq.~(\ref{Etot}).  
%\clearpage

Figure~\ref{fig05} shows the comparison of the $\Delta$ values for protons and neutrons in $^{220}$ Rn obtained by minimizing the total energy given by (\ref{Etot}) in nine-dimensional space over 
$\alpha_{20}$, 
$\alpha_{22}$, $\alpha_{30}$, $\alpha_{31}$, $\alpha_{32}$, $\alpha_{33}$, $\alpha_{40}$ and $\Delta_n$, $\Delta_p$. The $J_{\perp}$ moment of inertia is also presented. For a low value of spin ($I=2~\hbar$), the proton $\Delta_{p}$ decreases with increasing deformation $\alpha_{30}$, reducing the pairing energy correction, thus effectively increasing the contribution of the Strutinsky shell energy correction, capable of creating an octupole energy minimum. As seen, this trend is undergoing a kind of saturation, starting from $I\approx 5~\hbar$.
At the same time the neutron $\Delta_{n}$ changes rather weakly with the deformation $\alpha_{30}$. This behavior for $\Delta_{n}$ persists also for higher spins. 

Comparison of the pairing $\Delta$s as a function of spin and axial octupole deformation for the studied nuclei $^{220}$Rn and $^{226}$Ra are shown in Fig.~\ref{delty}. 
At first glance, one can notice a rapid decrease in the neutron pairing parameter $\Delta$ with increasing spin
throughout the entire range of octupole axial deformation that, in the case of $^{220}$ Rn, practically falls to
$\Delta_{n}\approx 0$ for $I=9~\hbar$ and $\alpha_{30}\approx 0.2$ (region of strongest shell effects), indicating the transition from superfluid to normal state in the neutron distribution. 

A less pronounced effect of this kind has been noted in $^{226}$Ra, but some fluctuations of $\Delta_{n}$ are observed for low spins, $I=0,1$, implies a slight reorganization in neutron shell structure, reflected in Fig.~(\ref{fig01}) as a minor decrease in the depth of the potential energy well just after the collective rotation is switched on.

In contrast, the proton $\Delta$ field is poorly affected by the external rotation in both nuclei, because the pairing correlations are very weak for I=0. 
The above observations for $\Delta_{n}$ can be explained by the fact that, as the angular velocity of rotation increases, the interacting neutron valence levels in $^{220}$Rn and $^{226}$Ra (panels a-b) under the influence of the Coriolis force evolve rapidly towards maximum alignment, leading to decoupling of the corresponding time-reversal correlated pairs. Nevertheless, these arguments do not entirely apply to proton interacting orbitals in both nuclei (panels b-d), which, due to the less favorable values of the quantum numbers $j$ and $m$, are much less susceptible to alignment.
These theoretical results are in line with the predictions on the octupolarity character of our two nuclei, contained in the experimental work \cite{cocks}.

\subsection{Impact of rotation term}
  
\begin{figure}[!bt]
\includegraphics[width=2.70in]{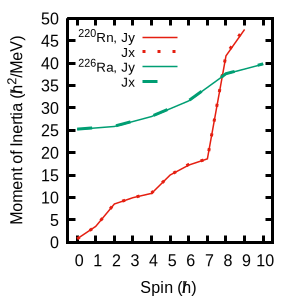}
\caption{\label{inertia}  
Moment of inertia for in the equilibrium points of $^{220}$Rn and $^{226}$Ra as function of spin. 
}
\end{figure}

Knowledge of microscopic moments of inertia as a function of spin (Fig.~\ref{inertia}) becomes
particularly important for nuclei, where the character of octupole instabilities can dynamically vary with rotation frequency.
Then, most often, the nucleus is not an easily predictable classical rotor of spin $I$ with energies proportional to $I(I+1)$, but its shell structure and consequently, its equilibrium shape may experience strong variations with collective rotation. 
In such nuclei, there may happen a change in the nature of the octupole instability: an initially vibrational nucleus may gradually become octupole-deformed, where, with increasing rotation frequency, such an energy well stabilizes its minimum position and depth. Such a phenomenon is favored by substantial ''softness'' of the potential energy function to changes in octupole deformation. In such a case, the equilibrium configurations corresponding to neighboring spin values within a band may differ significantly in terms of deformation and pairing interaction properties defined, e.g. by the $\Delta$ gap parameter. As shown later, this behavior is fully observed in the nucleus $^{220}$Rn in contrast to $^{226}$Ra, whose intrinsic structure is less sensitive to the influence of external rotation.

Studying Fig.~\ref{inertia} we are convinced that the use of constant, rigid-body or even microscopic moments of inertia across the energy band would be meaningless in nuclei with an internal structure highly susceptible to external rotation, such as $^{220}$Rn nucleus. On the other hand, in the second $^{226}$Ra system studied here, the low-energy sequences resemble the behavior of a classical rotor. For this case, a comparison of model bands generated using a constant moment of inertia with the one in which it changes with spin can reveal the level of non-collectivity in the rotational mode.
\begin{figure}[H]
\includegraphics[width=3.0in]{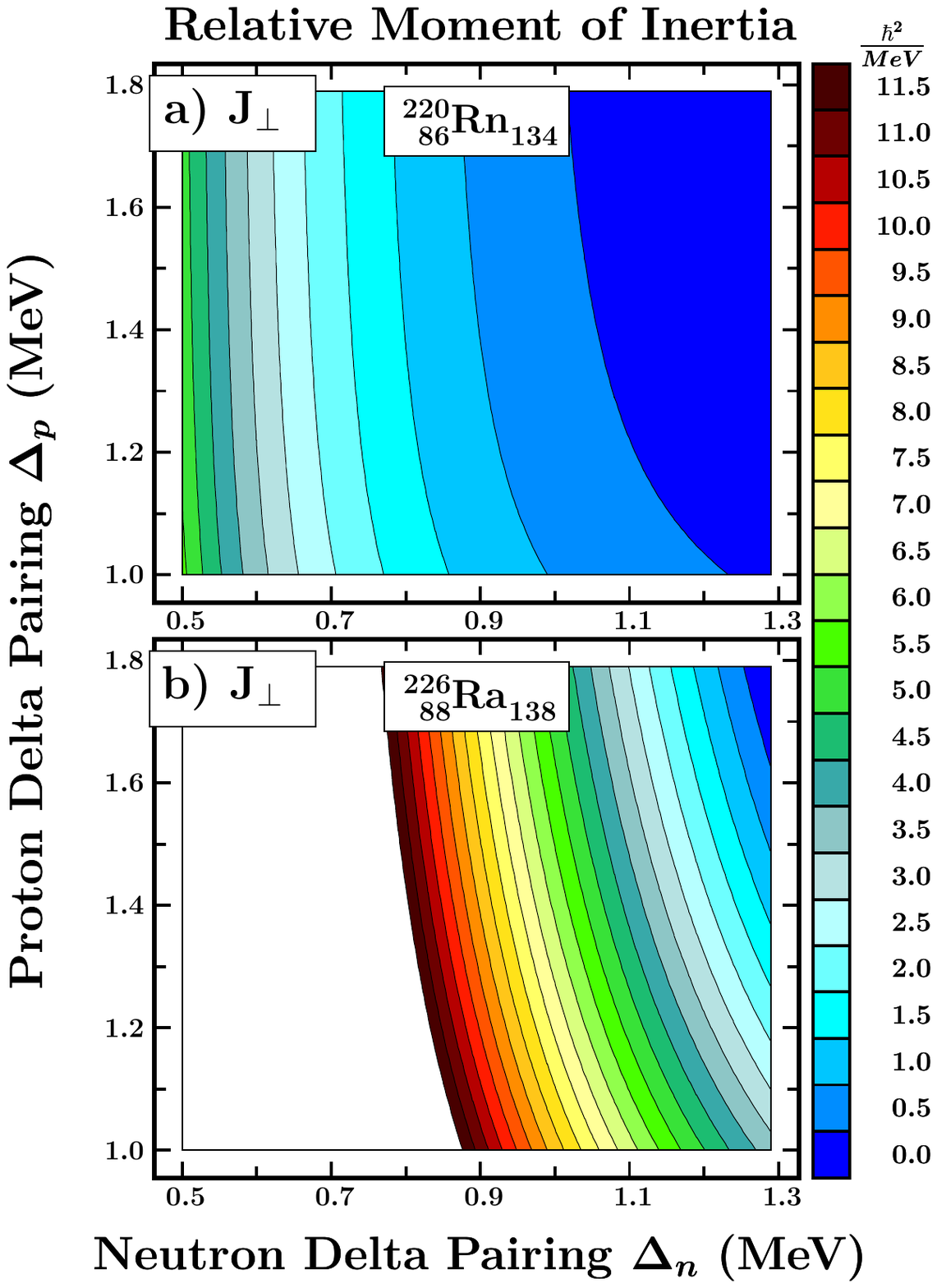}
\caption{\label{dndpJ}  
Changes of moment of inertia for the minimum-energy deformation with respect to its ground-state value in $^{220}$Rn and $^{226}$Ra nuclei for various pairing $\Delta_n$ and $\Delta_p$ parameters. 
}
\end{figure}
The pairing $\Delta_{n,p}$ is known to modify the values of the moment of inertia as shown in (Fig.~\ref{dndpJ}). 
Especially, in the limit of $\Delta \to 0$ this moment can increase even by a factor of 2-3 compared to the value in superfluid equilibrium, reaching in a well-deformed quadrupole nucleus the value of the moment of inertia of a rigid body.
The plotted variation of the ground-state microscopic moment of inertia as a function of pairing $\Delta_n$ and $\Delta_p$ gaps evolves smoothly but the rate of increase depends on the changes of occupation probabilities with rotation speed, different in both nuclei.
In well deformed $^{226}$Ra ($\alpha_{20}\approx 0.17,\alpha_{30}\approx 0.1$)
the moment of inertia $J_{\perp}$ grows much faster with decreasing pairing $\Delta$ than in nearly spherical $^{220}$Rn. This study sheds some light on the problem of the {\it pairing collective vibration} effect, to which we also refer in the following sections, results in a value for the most probable pairing $\Delta$ that is typically lower by 20-30$\%$ on average than its static BCS solution. 

This implies that for the range of $\Delta_n$, $0.5\leq\Delta_n\leq 0.8$, the changes of moments of inertia are the fastest. Such $\Delta_n$ values occur for spins $I=4-7\hbar$, where the transition energies between the $4^+$ and $6^+$ or $3^-$ and $5^-$ states, reported in the following sections, are visibly overestimated. 
Therefore, a thorough study of the pairing vibration problem, gives hope for obtaining larger moments of inertia in this range of spins than at present, which would significantly reduce the mentioned transition energies. A similar reasoning leads us to the general conclusion that this effect will cause the expected ''compression'' of both discussed collective spectra in $^{226}$Ra in a rather homogeneous manner.

Changes in the moment of inertia with spin $I$ obtained from the minimization of the energy (Fig.~\ref{Etot}) over nine degrees of freedom of deformation and pairing fields: $\alpha_{20}$, $\alpha_{22}$, $\alpha_{3\nu}$, ($\nu=0,1,2,3$), $\alpha_{40}$, $\Delta_n$, $\Delta_p$, denoting axial and non-axial quadrupole and octupole deformations, and neutron and proton pairing gaps, respectively, is presented in Fig.~\ref{inertia}. As both discussed here nuclei are practically axially symmetric in their energy equilibrium states, the $J_{x}$ and $J_{y}$ values are almost indistinguishable.  

\begin{figure}[!bt]
\includegraphics[width=3.0in,angle=0]{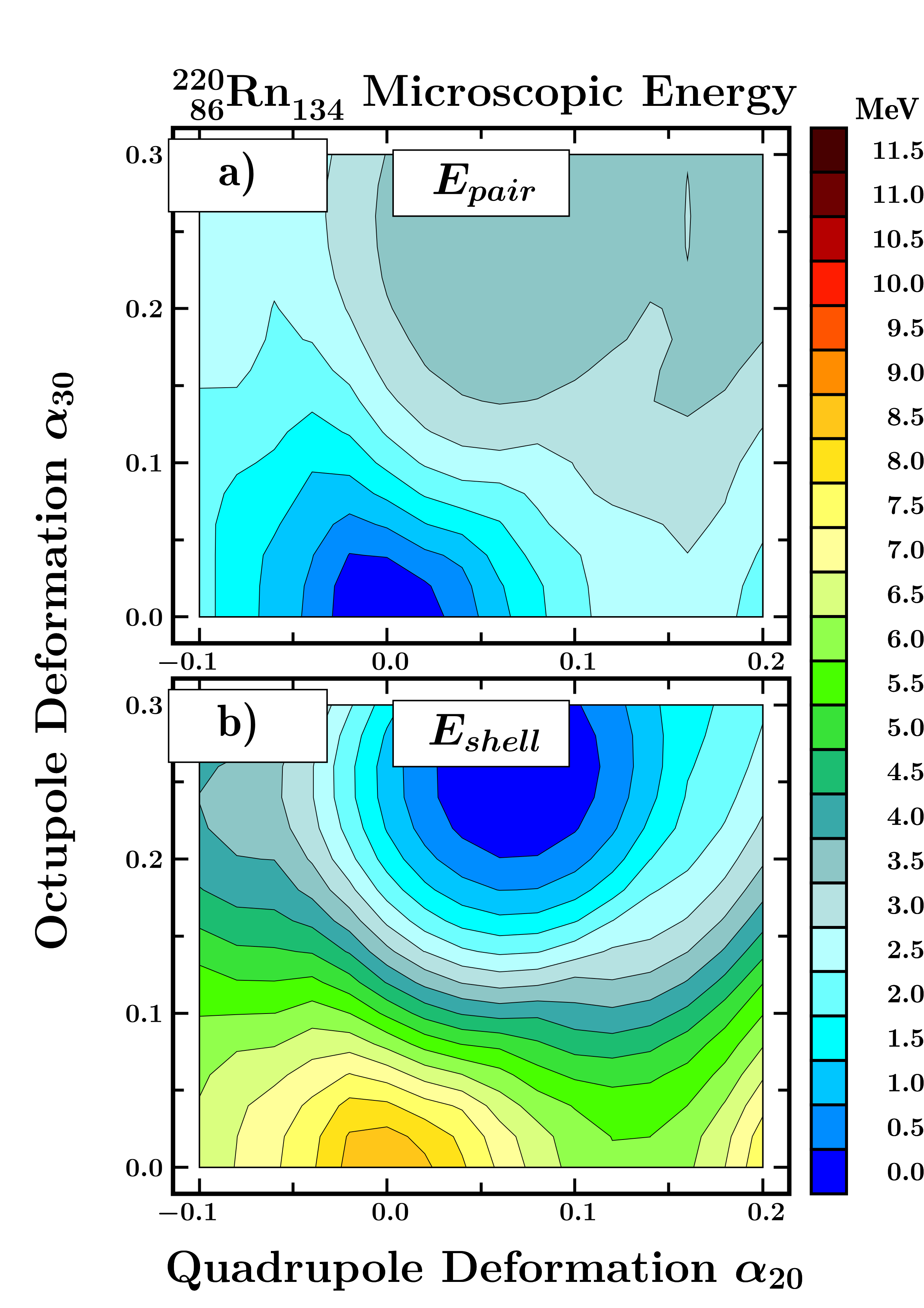}
\caption{\label{micro} The microscopic (pairing and shell) part of total energy in $^{220}$Rn for $I=0$, minimized over axial and non-axial quadrupole $\alpha_{2\mu}$ and octupole $\alpha_{3\mu}$ deformation parameters, displayed in ($\alpha_{20},\alpha_{30}$) deformation plane.
}
\end{figure}
An interesting trend of the moment of inertia can be observed in $^{220}$Rn. As shown, in case of no rotation, this nucleus is in the ground state almost spherical with $\alpha_{20}\approx 0.05$ and $\alpha_{30}=0$. In view of earlier remarks, the collective rotation of such a configuration cannot be
described by rigid body moments of inertia that give $J_x$ a few dozens of $\hbar^2/MeV$ but rather with the use of microscopic estimates, producing a realistic, small value of $J_x\approx 2-3\,\hbar^2/MeV$. As the rotation angular velocity increases, that is, for $I>2\hbar$, the total energy minimum starts to gradually shift toward octupole deformation $\alpha_{30}\approx 0.15$, which has not yet been achieved for $I\geq 5$. In parallel, the emerging reflection-asymmetric energy well is becoming deeper and deeper, stabilizing the octupole permanent deformation in this nucleus. 
Such phenomenon is a result of relatively weak quantal effects in this nucleus. 
In terms of the macroscopic-microscopic approach, the pairing energy correction occurs to be strongly sensitive on collective rotation, particularly in the neutron part, as depicted in Fig.~\ref{delty} by fast lowering of the $\Delta_n$ parameter with $I$, thus tending to weaken the total pairing contribution in the region of emerging octupole energy ''pocket''. On the other hand, the spin-independent Strutinsky shell energy term, whose value is, in general, greater than the pairing correction energy
(see, Fig.~\ref{micro}) and having opposite sign, is effectively responsible for substantial susceptibility of $^{220}$Rn to octupole deformation (octupole softness). As a result, when the pairing term decreases with increasing spin, the shell energy becomes more and more dominant, creating an increasingly pronounced octupole energy minimum. 

As already mentioned, intrinsic structure rearrangements in nuclei under external rotation are the Coriolis and centrifugal effects which result from the coupling of collective rotation to the single-particle angular momenta of the valence nucleons, forcing them to align along the angular velocity vector, see, e.g. \cite{devoight-dudek}. This alignment obviously leads to a reduction in the total energy of the rotating system. The orbitals most susceptible to this type of interaction are those with a relatively large total angular momentum and the minimum possible projection of this momentum onto the quantization OZ axis (as parallel as possible to the collective rotation axis). 

This mechanism contributes to a gradual weakening of the pairing field with increasing spin until one or a few nucleonic Cooper pairs break completely, thus affecting a significant increase of the moment of inertia. The above microscopic mechanism can be easily explained in terms of the sign of the derivative of a single-particle orbital $(\epsilon_{\nu},\nu^{\omega})$ over rotational frequency $\omega$, $d\epsilon_{\nu}/d\omega$, given as e.g. in \cite{devoight-dudek}
\begin{equation}
\frac{d\epsilon_{\nu}}{d\omega}=-\hbar \langle \nu^{\omega}|\hat j_{\perp}|\nu^{\omega}\rangle.
\label{dedomega}\end{equation}
Leaving aside the question of the consequences of breaking the time-reversal symmetry in mean-field Hamiltonian due to the external rotation, relation (\ref{dedomega}) shows basically that with increasing $\omega$, the orbitals above Fermi surface, for which 
$\langle \nu^{\omega}|\hat j_{\perp}|\nu^{\omega}\rangle$ is non-negligible and negative, can fall down closer to Fermi surface enhancing the probability of scattering a correlated pair from under Fermi level. This case is in favor of rapid alignment. Otherwise, if such orbitals are moving away from the Fermi surface, the correlated pairs respond more weakly to increasing rotation speed, tending to reduce the pair-breaking probability and the back-bending effect in moments of inertia.

Another important effect that can modify the moments of inertia is vibration of the pairing field, mentioned as collective pairing vibrations, proposed in \cite{Bes} and successfully applied, e.g. in \cite{zajac,prochniak,rohozinski} to reproduce and predict the spectroscopic observables using the Bohr Hamiltonian model. Briefly, apart from vibrations and rotations of nuclear surface described by the change in time of the deformation tensor components $\alpha_{\lambda\mu}$ and Euler angles, a nucleus may rotate and vibrate in the space of the pairing $\Delta$ and so called $\theta$ {\it gauge} angle parameters, considered as additional collective variables. Similarly as in the case of our Hamiltonian (\ref{Hcoll}), the collective rotation in $\theta$ space and vibration in $\Delta$'s can be considered independently with some specific inertia parameters. To a fair approximation, the coupling of this motion to the surface vibrations and rotations is negligible.

The resulting pairing $\Delta$ parameters are essentially lower by about 20-30$\%$ than the static BCS-PNP solutions, thus providing some 30-60$\%$ increase in microscopic moments of inertia (\ref{crankingMoI}). Since this effect is not considered in detail as part of this work, we assume that, on average, all moments of inertia in (\ref{Hcoll}) will be increased by a factor of 1.4, improving the consistency of rotational level energy estimates with experimental data. However, we are aware that above mentioned effect of pairing vibrations may change with spin of the nucleus, which will make the multiplying factor also depend on it. We will soon devote a separate work to this issue.

In turn, centrifugal effects, analogous to the classical behavior of an elastic body undergoing rotation with respect to a given axis perpendicular to the nucleus symmetry OZ axis, force changes in the elongation of the nucleus with respect to the symmetry axis, usually as a function of the rotation frequency with a concomitant narrowing of the cross-sectional area of the nucleus. This is referred to as the {\it stretching} effect \cite{pomorscy-quentin}. 
Such induced stretching, by changing the shape of the nucleus, leads to reorganizing its internal structure and thus also its potential energy and moments of inertia as direct functions of the deformation and pairing variables.  

As can be deduced, the primary goal of present calculations is first of all, to accurately determine the microscopic moments of inertia of the nucleus in the deformation and the pairing variables space at fixed values of the nucleus spin and then use them to determine the energy and structure of the rotational levels as eigen-solutions of Hamiltonian (\ref{Hcoll}). Its eigenfunctions are further used to calculate the reduced transition probabilities B(E$\lambda$) to model the low-lying bands.  

%%%%%%%%%%%%%%%%%%%%%%%%%%%%%%%%%%%%%%%%%%%%%%%%%%%%%%
\subsection{Electric transition probabilities}
%%%%%%%%%%%%%%%%%%%%%%%%%%%%%%%%%%%%%%%%%%%%%%%%%%%%%%%
Convincing evidence that most nuclei have non-zero reflection-symmetric quadrupole deformation comes from measurements of B(E2) electric transitions resulting from the charge distribution in the quadrupole deformed body. Similarly, octupole-shaped proton distribution contribute to enhanced B(E1) electric dipole and B(E3) octupole transitions between rotational states of opposite parity. 

It is known, however, that the B(E1) value consists not only of the contribution of purely collective effects, but also of microscopic shell effects. The role of B(E1) rates to unambiguously resolve the nature of octupole instability in a given nucleus is therefore limited.  The collective component of the dipole transition operator is the result of the interplay of the purely dipole deformation variables, which, if small enough, contribute essentially to the displacement of the nucleus' center of mass relative to the onset of the coordinate system, and the magnitude of the coupling between octupole and quadrupole deformations. Such determined, say, macroscopic values of B(E1) matrix elements are still significantly influenced by the aforementioned fluctuating single-particle effects, as well as by the inhomogeneous electric charge distribution associated with the variable curvature of the nucleus surface, as discussed in \cite{butler}, which, however, are not considered in this work.

A more reliable indicator of octupole correlations may therefore be B(E3) transitions, which
result directly from the reflection-asymmetric charge distribution in the nucleus volume, which
essentially depends on the collective behavior of the nucleus. Even if both octupole oscillations and stable octupole deformations contribute together to the B(E3) value, it is possible to distinguish between the two types of octupole instability by performing comparative analyses with other available observables, such as energy spacing between band states varying with spin, values of experimental spin alignment coefficients as a function of rotational frequency, branching coefficients of B(E$\lambda$)s, etc.

\begin{figure}[!bt]
\setlength{\unitlength}{0.1\textwidth}
\begin{picture}(4,8.)
%\put(2,9){a)} 
\put(-0.4,4.){\includegraphics[width=2.5in]{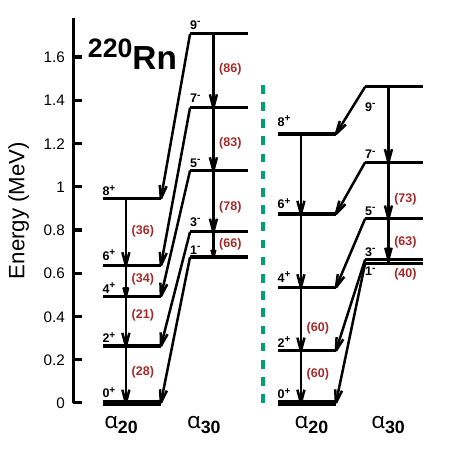}}
\put(0.,7.7){\bf \large a)          $J(\Delta_{n,p}$;I)    $\quad$  experiment} 
    \put(-0.4,0){\includegraphics[width=3.4in]{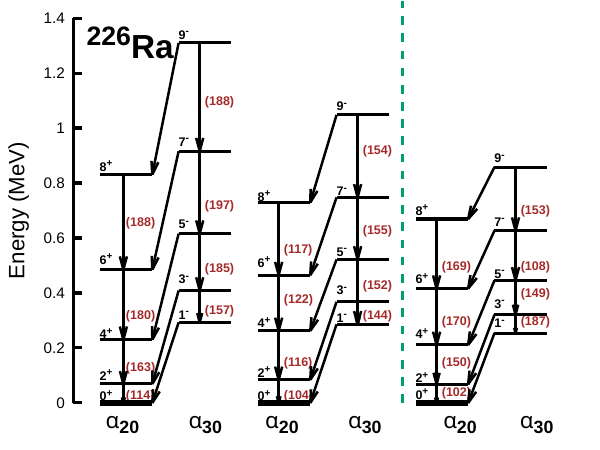}}
\put(0.,3.6){\bf \large b) $\quad$     $J_0$   $\quad$        $J(\Delta_{n,p}$;I)  $\quad$       experiment}
\end{picture}
\caption{\label{spectra} Comparison of theoretical and experimental \cite{butler} bands for $^{220}$Rn (a) and $^{226}$Ra (b) obtained with the spin-dependent moment of inertia $J(\Delta_{n,p};I)$ and, in case of $^{226}$Ra, additionally with constant $J_0$ value. Transition probability B(E2)(efm) are marked in brackets.
}
\end{figure}

The experimental data for $^{220}$Rn and $^{226}$Ra include not only the energy spectra, but also selected values of the reduced transition probabilities B(E1), B(E2) and B(E3). A comparison of the measured and theoretical energy spectra and transitions is shown in Fig.~\ref{spectra}, where $J_0$ denotes the band-wide constant moment of inertia calculated for $I=0$ and static BCS-PNP values of the pairing gaps, 
$\Delta^{(0)}_{p,n}$, while $J(\Delta_{p,n};I)$ refers to its spin-dependent value.
Apart from B(E2)s, also dipole and octupole transition probabilities have been extracted from measurement. The comparison with theoretical predictions are given in Tabs.~\ref{tab_e1} and \ref{tab_e3}.
\begin{table}[H]
\caption{Transitions B(E1) and B(E3) for $^{220}$Rn.}
\begin{center}
   \begin{tabular}{ ccccc }
 \hline 
&B(E1) (W.u.$\cdot10^{-5}$)&  $J(\Delta_{n,p};I)$ & exp. \cite{nature} \\
\hline
&$1^- \rightarrow 0^+$ & 15&35\\
&$3^- \rightarrow 2^+$ &90&$<200$\\
&$5^- \rightarrow 4^+$ &45&3$^{+2}_{-1.6}$\\
&$7^- \rightarrow 6^+$ &200&$<5000$\\
&$9^- \rightarrow 8^+$ &222&\\
\hline
&B(E3) (W.u.)&  $J(\Delta_{n,p};I)$ & exp. \cite{nature} \\
\hline 
&$3^- \rightarrow 0^+$ &26& 33$\pm 4$\\
&$5^- \rightarrow 2^+$ &21&90$\pm50$\\
&$7^- \rightarrow 4^+$ &6.4&\\
&$9^- \rightarrow 6^+$ &9.8&\\
\hline
\end{tabular} \label{tab_e1}
\end{center}
\end{table}
%%%%%%%%%%%%%%%%%%%%%%%%%%%%%%%%%%%%%%%%%%%%%%%%%%%%
\begin{table}[H]
\caption{Transitions B(E1) and B(E3) for $^{226}$Ra.}
\begin{center}
   \begin{tabular}{ ccccc } 
\hline 
&B(E1) (W.u.$\cdot10^{-5}$)& $J_0$ & $J(\Delta_{n,p};I)$ & exp. \cite{ wolersheim} \\
\hline
&$1^- \rightarrow 0^+$ &100&88&35$\pm 13$\\
&$3^- \rightarrow 2^+$ &129&107&22$\pm 5$\\
&$5^- \rightarrow 4^+$ &137&105&33$\pm 5$\\
&$7^- \rightarrow 6^+$ &140&101&69$\pm 9$\\
&$9^- \rightarrow 8^+$ &142&96&127$\pm 21$\\
\hline 
&B(E3) (W.u.)& $J_0$ & $J(\Delta_{n,p};I)$ & exp. \cite{ wolersheim} \\
\hline 
&$3^- \rightarrow 0^+$ &0.007&13& 0.157$\pm 0.015$\\
&$5^- \rightarrow 2^+$ &0.009&16.2&\\
&$7^- \rightarrow 4^+$ &0.006&16&\\
&$9^- \rightarrow 6^+$ &0.002&7.5&\\
\hline
\end{tabular}\label{tab_e3}
\end{center}
\end{table}
Fig.~\ref{spectra} gives the comparison of rotational levels of the ground-state and low energy negative-parity bands, generated by our collective Hamiltonian with microscopic variable mass tensor and spin-dependent moments of inertia, with experimental data. From more than a dozen eigensolutions of the Hamiltonian characterized by energy, spin and phonon number of either quadrupole or octupole excitation (band head), we selected several that were identified as measured states. The theoretical bands are made up of low-lying energy states of a given parity, which are connected by the strongest intraband transitions B(E2), which, as can be seen, have been shown to be comparable to the measured ones. 
Both positive and negative parity bands must be connected by a B(E1) dipole of an order similar to that given in the experiment. The decisive test of whether a given set of states constitutes the desired bandwidth is the values of the interband octupole transitions B(E3), which are assumed to differ from the measured ones up to at most 2 orders of magnitude. We are aware that better theoretical estimates of that observables exist, e.g. \cite{budaca}. Nevertheless, our aim is to present the results obtained in an approach in which there are no parameters adapted to the experimental data relating directly to energy levels or electromagnetic transitions.

As can be seen, in the nucleus $^{220}$Rn, the discordance of the estimation of the energies of low-lying states obtained with spin-dependent moments of inertia for spins from $0^+$ to $8^+(9^-)$, compared to experimental values, varies from a several tens of keV in the entire GS band with the aforementioned exception of about 200 keV for a single $E_{6^+}-E_{4^+}$ transition energy. Similarly,
an average discrepancy in the negative-parity band is less than 100 keV. In particular cases, it systematically grows with spin from 40 keV between $1^-$ partners, reaching about 180 keV for the pair of $9^-$ states. \\
 A similar conclusions apply to $^{226}$Ra, where the maximum inaccuracy within the range of 100-200 keV is noted between the theoretical and experimental $7^-$ and $9^-$ states. Comparing the bands obtained using the spin-dependent moments of inertia $J(\Delta_{n,p};I)$ and their counterparts obtained with the spin-independent $J_0$ moment, we find that the latter sequences are generally increasingly stretched in energy with growing spin than the former, proving that the nature of this nucleus is far from the collective behavior of $I(I+1)$, but to a lesser extent than in $^{220}$Rn.\\
The fact that the discrepancy in question increases as a function of spin indicates the need for a more accurate treatment of the pairing forces in microscopic moments of inertia, in particular considering the spin-dependent effect of pairing vibrations, taken into account only on average by increasing {\bf all} moments by 40$\%$.
We believe that this effect will help us to reduce the extremely small (as little as 10 keV) energy gap between $1^-$ and $3^-$ measured in $^{220}$Rn, since it is supposed to result from the relatively rapid change of shape, and thus moment of inertia (from $2.3\hbar^2/MeV$ for the $I=1^-$ to above $10\hbar^2/MeV$ for $3^-$) as the nucleus alters its octupolarity character from initially vibrational, gradually toward a stably deformed octupole with 
$\alpha_{30}\approx 0.1$ and $\alpha_{20}\approx 0.05$ for $I=3^-$.
This relatively rapid increase in moment of inertia with spin makes both model spectra in question, generated at constant moment of inertia $J_0<1\,\hbar^2/MeV$ would be highly stretched in energy and thus not comparable in any way to the experimental.\\
As for the electromagnetic transition probabilities, highlighted in Fig.~\ref{spectra} quadrupole B(E2) and listed in Tabs.~\ref{tab_e1} and \ref{tab_e3} B(E1) and B(E3), they remain in general agreement to an order of magnitude with the measured values in both nuclei. The difference of two orders of magnitude in 
B(E3, $3^-\to 0^+$)=13 W.u. (versus experimental 0.157 W.u., see Tab.~\ref{tab_e3} for $^{226}$Ra), with respect to the experimental value is, first, due to the absence of couplings of the type 
$\alpha_{30}\times\alpha_{20}$ in the octupole transition operator, assumed in this work to be 
$\hat Q_3=3ZR_0\,\alpha_{30}/4\pi$ and second, as a result of above mentioned uncertainty about the intrinsic structure of the theoretical $3^-$ state, which ranks too high in energy relative to $1^-$. The last statement seems obvious if one compares the two estimates of B(E3, $3^\to 0^+)$ in Tab.~\ref{tab_e3}, which differ by more than three orders of magnitude for two different values of the applied moment of inertia.
In turn, the second-order term in $\hat Q_3$ can have a significant impact in a well-deformed state, where both the equilibrium octupole and quadrupole $\alpha-$type deformations have significant values, of 0.11 and 0.17, respectively. Note that the corresponding value of B(E3) in the on-average very weakly quadrupole-deformed $^{220}$Rn, where the product of $\alpha_{30}\times\alpha_{20}$ is neglected, is quite well reproduced using the above simplest form of $\hat Q_3$.\\
The general agreement of the results presented above with the measured data proves the realistic nature of the proposed collective Hamiltonian model, as well as the correctness of the choice of states representing the experimental bands. As can be seen from the aforementioned inconsistencies, the concept of pairing collective oscillations still needs a proper implementation.
Since the model actually has only one free parameter in the form of a constant coefficient, rescaling the moments of inertia due to that effect, we believe that the obtained estimates are satisfactory.

%%%%%%%%%%%%%%%%%%%%%%%%%%%%%%%%%%%%%%%%%%%%%%%%%%
\subsection{Conclusions}\label{wnioski}
The presented work is devoted to investigate the character of the octupole instabilities in two nuclei, $^{220}$Rn and $^{226}$Ra, which were predicted by the experimentalists more than two decades ago.
Since these nuclei are not classical rotors, the effects of Coriolis and centrifugal effects are proved to modify the single-particle structure around the Fermi levels. As shown, even if the mass numbers in these nuclei
differ by six mass units (2 protons and 4 neutrons), the response of the pairing field on external rotation leads to significantly different ways of evolution of potential energy wells towards octupole deformation with increasing spin. 
These effects are also evident in the structure of the level diagrams of both nuclei, where significant deviations from the parabolic dependence of energy on spin are observed.

Although, the model used here is not a self-consistent one, involving a modern effective nucleon-nucleon interaction and enabling for a proper treatment of subtle microscopic phenomena, a complex transition from vibrational to stable deformed configuration is ultimately achieved. This is possible due to large precision of the calculations made on very dense deformation and pairing $\Delta$ grid and ingredients of the macroscopic-microscopic model well tested in different fields of nuclear physics.
The collective model used to determine the rotational states
with spin-dependent moments of inertia is capable of reproducing on average the energies as well as the electric transitions within order of magnitude accuracy.

The problem of collective vibrations of the pairing field is implemented only on average by rescaling the microscopic moments of inertia by a constant for all states factor of $1.4$ to preserve the mutual layout of the positive and negative parity band members. 
A more detailed study of this very important and also complex phenomenon as a function of nuclear spin will be the subject of future research.
%%%%%%%%%%%%%%%%%%%% Figure 3 begin %%


\begin{thebibliography}{999}
\bibitem{butler} P.A. Butler, W. Nazarewicz, Rev. Mod. Phys. {\bf 68}, 349 (1996).
\bibitem{cocks} J.F.C. Cocks, et al. Nucl. Phys. A {\bf 645}, 61 (1999).
\bibitem{Dudek-Dedes1} J. Dudek, I. Dedes, A. Baran et al. Eur. Phys. J. Spec. Top. 233, 965 (2024). 
\bibitem{Dudek-Dedes2} J. Dudek, D. Curien, I. Dedes, K. Mazurek, S. Tagami, Y. R. Shimizu,
                    and T. Bhattacharjee, Phys. Rev. C {\bf 97}, 021302(R)(2018).
\bibitem{nadirbekov}
M. S. Nadirbekov, O. A. Bozarov, S. N. Kudiratov, and N. Minkov, Int. Jour. of Mod. Phys. E {\bf 8}, 2250078 (2022). 

\bibitem{proxy} A. Restrepo, J. P. Valencia, Phys. Rev. C {\bf 110}, 054312 (2024).
\bibitem{bonatsos} D. Bonatsos, A. Martinou, S.K. Peroulis, T.J. Mertzimekis, N. Minkov, 
Atoms {\bf 11}, 117 (2023). 
\bibitem{AD-KM-AG-2016} A. Dobrowolski, K. Mazurek, A. G\'o\'zd\'z,
Phys. Rev. C {\bf 94}, 054322 (2016).
\bibitem{AD-KM-AG-2018} A. Dobrowolski, K. Mazurek, A. G\'o\'zd\'z,
Phys. Rev. C {\bf 97}, 024321 (2018).
\bibitem{AD-AG-KM-2018} A. Dobrowolski, A. G\'o\'zd\'z, K. Mazurek,
Acta Phys. Polonica. {\bf B 48}, 565-572 (2017).
\bibitem{BohrHam} A. Bohr, Mat. Fys. Medd. Dan. Vid. Selsk. {\bf 26}, 14 (1952).
\bibitem{Inglis} D. R. Inglis, Phys. Rev. {\bf 96}, 1059 (1954); {\bf 103}, 1786 (1956); S. T. Balyeav, Nucl. Phys. {\bf 24}, 322 (1961).
\bibitem{crank} K. Pomorski, T. Kaniowska, A. Sobiczewski, S.G. Rohozinski,
Nucl. Phys. A {\bf 283}, 394 (1977).
\bibitem{myCPC2016} A. Dobrowolski, K. Pomorski, J. Bartel, Comp. Phys. Comm. {\bf 199}, 118 (2016).
\bibitem{colpair} A. G\'o\'zd\'z, K. Pomorski, M. Brack, E. Werner, Nucl. Phys. A {\bf 442}, 50 (1985); A. G\'o\'zd\'z, K. Pomorski, Nucl. Phys. A {\bf 451}, 1 (1986). 
\bibitem{stab} K. Pomorski, A. Dobrowolski, B. Nerlo-Pomorska, M. Warda, J. Bartel, Z.G. Xiao, Y.J. Chen, L.L. Liu, J-L. Tian, X.Y. Diao, Eur. Phys. Journ. A {\bf 58}, 77 (2022). 
\bibitem{pomorski1} T. Kaniowska, A. Sobiczewski, K. Pomorski, S.G. Rohozi\'nski, Nucl. Phys. A {\bf 274}, 151 (1976).
\bibitem{dudek80} J. Dudek, A. Majhofer and J. Skalski, J. Phys. G {\bf 6}, 447 (1981).
\bibitem{gen_Hrot} J. Dudek, A. G\'o\'zd\'z and D. Ros\l y, Acta Phys. Polonica B {\bf 32}, 2625 (2001).
\bibitem{gen_Hrot2} M. Mi\'skiewicz, A.~G\'o\'zd\'z, J. Dudek, Int. J. Mod. Phys. E {\bf 13}, 127 (2004).
%\bibitem{maj2} A. Maj, et al., Nucl. Phys. {\bf A 731}, 319 (2004).
\bibitem{Pomorski-Dudek-2003} K Pomorski, J Dudek, Phys. Rev. C {\bf 67}, 044316 (2003).
\bibitem{devoight-dudek} M.J.A. Devoigt, J. Dudek,  Rev. Mod. Phys. 55, 949-1064 (1983).
\bibitem{Bes} D. R. B `es, R. A. Broglia, R. P. J. Perazzo, and K. Kumar, Nucl. Phys. A {\bf 143}, 1 (1970).
\bibitem{zajac} K. Zajac, L. Prochniak, K. Pomorski, S.G. Rohozinski, and J. Srebrny,
Acta Phys. Polonica B {\bf 34}, 1789 (2003). 
\bibitem{prochniak} L. Pr\'ochniak, K. Zajac, K. Pomorski, S.G. Rohozinski, J. Srebrny,
Acta Phys. Polonica B {\bf 33}, 405 (2002). 
\bibitem{rohozinski} S.G. Rohozinski, K. Pomorski, L. Prochniak, K. Zajac, Ch. Droste, J. Srebrny, 
 Phys. At. Nucl. (Yad. Fiz.) 64, 1005 (2001).
\bibitem{pomorscy-quentin} B. Nerlo-Pomorska, K. Pomorski, P. Quentin and J. Bartel, 
Physica Scripta 89, 054004 (2014).
\bibitem{nature} L. Gaffney, P. Butler, M. Scheck, et al., Nature 497, 199 (2013). 
\bibitem{wolersheim} H.J. Wollersheim, et al Nucl. Phys. A {\bf 556}, 261 (1993).
\bibitem{budaca} R. Budaca, P. Buganu, A.I. Budaca, Phys. Rev. C {\bf 106}, 014311 (2022).

\end{thebibliography}
\end{document}